\documentclass[aps,prd,10pt,twocolumn,superscriptaddress,preprintnumbers]{revtex4}

\usepackage{bm}
\usepackage{braket}

\usepackage{blindtext}
\usepackage{epsfig, cancel}

\usepackage{latexsym}
\usepackage{natbib, comment}
\usepackage{mathrsfs,amsmath,amssymb,amsthm,amsfonts,tikz,graphicx,accents,hyperref,color}
\usepackage{url}
\usepackage{dcolumn}
\usepackage{multirow}
\usepackage{color}
\usepackage{cancel}
\usepackage{soul}
\usepackage[normalem]{ulem}
\usepackage{txfonts}
\usepackage{epsfig}
\usepackage{psfrag}
\usepackage{subfigure}
\hypersetup{colorlinks=true}
\usepackage{mathtools}
\usepackage{enumitem}
\usepackage{float}

\def\H0{{\text{H}\hspace*{-2.05mm}\text{H} 0\hspace*{-1.35mm}0\ }}

\usepackage[all]{xy}
\usepackage{slashed}
\usepackage{slashed,ccaption}
\usepackage{multirow}

\usepackage{caption}

\usepackage{array}
\hypersetup{ linktoc=all,
    colorlinks, linkcolor={palatinateblue},
    citecolor={brightpink}, urlcolor={amaranth}
}

\graphicspath{{Images/}}

\renewcommand{\d}[1]{\ensuremath{\operatorname{d}\!{#1}}}

\DeclareSymbolFont{extraup}{U}{zavm}{m}{n}
\DeclareMathSymbol{\varheart}{\mathalpha}{extraup}{86}
\DeclareMathSymbol{\vardiamond}{\mathalpha}{extraup}{87}
\makeatletter
\renewcommand*{\@fnsymbol}[1]{\ensuremath{\ifcase#1\or \clubsuit \or \vardiamond \or \varheart\or
    \spadesuit\or \mathparagraph\or \|\or **\or \dagger\dagger
    \or \ddagger\ddagger \else\@ctrerr\fi}}
\makeatother

\definecolor{rosy}{RGB}{230,235,252}
\definecolor{myframetitle}{RGB}{90,89,170}
\definecolor{myblocktitle}{RGB}{140,185,249}
\definecolor{mytitle}{RGB}{10,80,26}

\definecolor{darkgreen}{RGB}{27,130,45}
\definecolor{darkblue}{rgb}{0,0,0.3}
\definecolor{darkred}{rgb}{0.7,0,0}

\definecolor{light gray}{RGB}{220,220,220}
\definecolor{dark purple}{RGB}{108,0,217}
\definecolor{pink}{RGB}{190,20,100}
\definecolor{orang}{RGB}{193,63,0}
\definecolor{green}{RGB}{11,98,17}
\definecolor{darkpink}{RGB}{153,0,76}
\definecolor{bluegreen}{RGB}{0,102,102}
\definecolor{greenlagan}{RGB}{0,102,0}
\definecolor{redgreen}{RGB}{102,102,0}
\definecolor{Redgreen}{RGB}{153,76,0}
\definecolor{vividviolet}{rgb}{0.62, 0.0, 1.0}
\definecolor{amaranth}{rgb}{0.9, 0.17, 0.31}
\definecolor{palatinateblue}{rgb}{0.15, 0.23, 0.89}
\definecolor{brightpink}{rgb}{1.0, 0.0, 0.5}
\definecolor{cornflowerblue}{rgb}{0.39, 0.58, 0.93}
\definecolor{deepcarminepink}{rgb}{0.94, 0.19, 0.22}
\definecolor{radicalred}{rgb}{1.0, 0.21, 0.37}

\newcommand{\bTh}{\boldsymbol{\Theta }}

\newcommand{\bO}{\boldsymbol{\Omega}}

\usepackage[most]{tcolorbox}

\tcbset{highlight math style={left=02mm,right=02mm,top=02mm,bottom=02mm}} 
\usepackage{empheq}

\DeclareFontFamily{OT1}{rsfs}{}

\DeclareFontShape{OT1}{rsfs}{m}{n}{ <-7> rsfs5 <7-10> rsfs7 <10->rsfs10}{} 

\DeclareMathAlphabet{\mycal}{OT1}{rsfs}{m}{n}

\newcommand{\be}{\begin{equation}}
\newcommand{\ee}{\end{equation}}

\begin{document}



\title{Darboux Soft Hair in 3D Asymptotically Flat Spacetimes}
\author{Vahid~Taghiloo}\email{v.taghiloo@iasbs.ac.ir}
\affiliation{School of Physics, Institute for Research in Fundamental
Sciences (IPM), P.O.Box 19395-5531, Tehran, Iran}
\affiliation{Department of Physics, Institute for Advanced Studies in Basic Sciences (IASBS),
P.O. Box 45137-66731, Zanjan, Iran}

\begin{abstract}
In this paper, we construct a \textit{fully} on-shell solution space for three-dimensional Einstein’s gravity in asymptotically flat spacetimes using a finite coordinate transformation. This space is parametrized by four unconstrained codimension-one functions that parameterize geometrical deformations of the null infinity cylinder, known as Darboux soft hair. The symplectic form of the theory in terms of these functions adopts a Darboux form at the corner, thereby yielding two copies of the Heisenberg algebra. Utilizing a series of field redefinitions of boundary charges, we construct various Lie algebras, including four Kac–Moody algebras, four Virasoro algebras, and two centrally extended BMS$_{3}$ algebras. Intriguingly, we show that the bulk theory possesses a well-defined action principle without the need for any boundary Lagrangian in the Darboux frame. Conversely, in the hydrodynamics frame, a well-defined action principle with the Dirichlet boundary condition results in two Schwarzian actions at null infinity.

\end{abstract}
\maketitle

\section{Introduction}
Einstein’s three-dimensional gravity is distinguished by the lack of local bulk degrees of freedom and its \textit{local} triviality \cite{Staruszkiewicz:1963zza, Deser:1983tn, Deser1984ThreeDimensionalCG}. In the three-dimensional flat case, this means all solutions derived from Einstein’s equations are locally flat. Despite this local triviality, its solution space has a rich global structure. The constant representatives of the solution space encompass a variety of topological solutions, including flat cosmologies, spinning particles, conical defects, and conical excesses \cite{Deser:1983tn, Barnich:2012aw, Bagchi:2012xr, Barnich:2012xq} (for a review see \cite{Compere:2019qed} and references therein). Moreover, its asymptotic symmetries, akin to the 4-dimensional case \cite{1962RSPSA.269...21B, Sachs:1962wk, PhysRev.128.2851}, undergo an infinite dimensional enhancement \cite{Ashtekar:1996cd, Barnich:2006av}, which is of particular relevance to our current study.

The asymptotic symmetries for null infinity in three-dimensional flat spacetimes, first discovered in \cite{Ashtekar:1996cd}, satisfy a BMS$_{3}$ algebra. Subsequently, \cite{Barnich:2006av} demonstrated that the surface charge algebra incorporates a non-trivial central extension term. Over the past decade, various extensions of these asymptotic symmetries have been explored, with one of the most general ones involving an isl$(2)_k$ algebra in \cite{Grumiller:2017sjh}.  Furthermore, a direct sum of  BMS$_{3}$ and a Heisenberg algebra has been found in \cite{Adami:2022ktn, Adami:2023fbm, Geiller:2021vpg}.

As implications of asymptotic symmetries, they serve as the initial step towards holography. The key insight is that these asymptotic symmetries represent the (gauge) symmetries of the boundary theory. For instance, the seminal work by Brown and Henneaux \cite{Brown:1986nw} demonstrates that with Dirichlet boundary conditions, the asymptotic symmetries of asymptotically AdS$_{3}$ spacetimes are characterized by two copies of the Virasoro algebra, which are the global symmetries of a two-dimensional conformal field theory. This finding propels us towards the AdS$_{3}$/CFT$_2$ correspondence \cite{Maldacena:1997re}. A similar approach based on the BMS symmetry group is currently being developed for flat space holography (for codimension-2 holography see \cite{Strominger:2017zoo, Pasterski:2021rjz, Raclariu:2021zjz} and references therein and also for codimension-1 holography see \cite{Bagchi:2022emh} and references therein).

In this paper, we present a fully on-shell solution space for three-dimensional Einstein’s gravity, incorporating four arbitrary functions at null infinity. As we deal with an unconstrained on-shell solution space, we can express the symplectic form of the theory in Darboux form. Consequently, we obtain two copies of the Heisenberg algebra as the surface charge algebra. 

A crucial point to highlight is that the charge algebra is not invariant under field redefinition. By employing a variety of field redefinitions,  known as the change of slicing technique \cite{Adami:2020ugu, Adami:2021sko, Adami:2021nnf, Adami:2022ktn, Taghiloo:2022kmh}, we are able to construct several charge algebras, including four copies of Kac–Moody algebras, four copies of Virasoro algebras, and two copies of centrally extended BMS$_{3}$ algebras. Interestingly, in the asymptotically flat case, we recover Virasoro algebras for the surface charge algebra, which are the canonical surface charge algebras for asymptotically AdS spacetimes \cite{Brown:1986nw}.

It is a well-established fact that a well-defined action principle in gravity necessitates the addition of a boundary action to the bulk action. For instance, in Einstein’s gravity with Dirichlet boundary conditions, the boundary term is commonly referred to as the Gibbons-Hawking-York term \cite{PhysRevD.15.2752, York:1972sj}. We will show that in the hydrodynamics frame with Dirichlet boundary conditions, the on-shell solution space yields two Schwarzian actions at null infinity (see \cite{Carlip:2016lnw, Bhattacharjee:2023sfd, Bhattacharjee:2022pcb} for other approaches for the appearance of Schwarzian actions at stretched horizon and null infinity. ). Conversely, in the Darboux frame, we find that there is no need for any boundary terms, and the pure Einstein bulk term suffices. 

The paper is organized as follows: Section \ref{sec:solution-phase-space-with-const} reviews the solution phase space for three-dimensional Einstein’s gravity \cite{Adami:2022ktn}. Section \ref{sec:solution-phase-space-without-const} constructs the fully on-shell solution space using finite coordinate transformations. Section \ref{sec:Darboux Basis} analyzes the symplectic structure, expressing the symplectic form in Darboux form. Section \ref{sec:Schwarzian-action} derives two Schwarzian actions from a well-defined action principle with Dirichlet boundary conditions. Section \ref{sec:Various Lie-algebra} reveals various Lie-algebras for surface charges using the change of slicing technique. Section \ref{sec:Outlook} provides a summary, discussion, and outlook. An appendix \ref{sec:isometry} explores the solution space’s isometries.
\section{Solution phase space with constraints}\label{sec:solution-phase-space-with-const}
In this section, we review the solution space for three-dimensional Einstein gravity constructed in \cite{Adami:2022ktn}.

We start from Einstein-Hilbert action in three dimensions without cosmological constant
\begin{equation}\label{action}
    S=\frac{1}{16\pi G}\int \d{}^3 x\, \sqrt{-g}\, R+\int \d{}^2 x_{\mu}\, \mathcal{L}_{\mathcal{B}}^{\mu}\, .
\end{equation}
The boundary Lagrangian $\mathcal{L}_{\mathcal{B}}^{\mu}$ is added to ensure a well-defined action principle, which will be refined later. The variation of this action results in the Einstein equation, $R_{\mu\nu}=0$.  For the construction of the solution space of this theory, we take the following ansatz for the line element 
\begin{equation}\label{metric}
    \d s^2= g_{\mu\nu}\d x^\mu \d x^\nu= -V \d v^2 + 2 \eta \d v \d r + {R}^2 \left( \d \phi + U \d v \right)^2\, ,
\end{equation}
where $v$, $r$, and $\phi$ respectively represent the null coordinate at (past) null infinity, the radial coordinate, and the periodic coordinate on the celestial circle. Apart from $\eta$ which is an arbitrary function independent of $r$, all other functions in the above ansatz are generic functions of all coordinates.

Three components of the Einstein field equation determine the radial dependence of these functions
\begin{subequations}\label{metric-components}
    \begin{align}
    {R}=& \Omega + r \, \eta \, \lambda \, ,\\
     U= &\ {\cal U} +  \frac{1}{\lambda \, {R}}\, \frac{ \partial_{\phi}\eta}{ \eta }  + \frac{{\Upsilon}-\Omega\, \partial_{\phi}\Pi}{2 \lambda { R}^2} \, ,\\
          V= & \ \frac{1}{\lambda^2}\Biggl[  -\mathcal{M}-2\, \text{Sch}[\sigma;\phi]+ \lambda\, \Omega\, (\partial_v\Pi-\mathcal{U}\partial_{\phi}\Pi)+ \left(\frac{\partial_{\phi}\eta}{\eta}\right)^2 \nonumber\\
          &\hspace{.5 cm}+\frac{({\Upsilon}-\Omega\, \partial_{\phi}\Pi)^2}{4  { R}^2} - \frac{2{R}}{\eta }\,    [\partial_v ( \eta \lambda )-\partial_{\phi}(\mathcal{U} \eta \lambda)] \nonumber\\
          &\hspace{.5 cm}+\left(\frac{{\Upsilon}-\Omega\, \partial_{\phi}\Pi}{{R}}\right) \frac{ \partial_{\phi}\eta}{ \eta } \Biggr] \, ,
    \end{align}
\end{subequations}
where 
\begin{equation}
    \Pi  :=2 \ln | \eta\, \lambda\, \Omega^{-1}|\, , \quad \sigma :=\int^{\phi}\d{}\phi'\lambda(v,\phi')\, ,
\end{equation}
and $\text{Sch}[X;\phi]$ denotes the Schwarzian derivative 
\begin{equation}
    \text{Sch}[X;\phi]  :=\frac{ X'''}{X'}-\frac{3}{2}\left(\frac{X''}{X'}\right)^2\, .
\end{equation}
The solution \eqref{metric-components} incorporates six arbitrary codimension one functions on null infinity: $\{\eta,  \mathcal{U}, \Omega, \lambda, \mathcal{M}, \Upsilon \}$. These are generic functions of $v$ and $\phi$ that emerge as integration constants once we integrate Einstein equations on $r$ to specify the $r$ dependence of the functions.  emerge as integration functions. They possess physical meaning, for instance, $ \mathcal{M}$ and $ \Upsilon$ represent the Bondi mass and Bondi angular momentum aspects, respectively.

The two remaining Einstein's equations impose constraints on these free functions
\begin{subequations}\label{M-Upsilon-EoM}
\begin{align}
    &\partial_{v}\mathcal{M}-\mathcal{U}\, \partial_{\phi}\mathcal{M}-2\mathcal{M}\, \partial_{\phi}\mathcal{U}+2\, \partial_{\phi}^3\mathcal{U}=0\, , \label{EOM-M}\\
    &\partial_{v}\Upsilon-\mathcal{U}\, \partial_{\phi}\Upsilon-2\Upsilon\, \partial_{\phi}\mathcal{U}-\lambda \partial_{\phi}\left(\frac{ \mathcal{M}}{\lambda^2}\right)+2\,\partial_{\phi}^3(\lambda^{-1})=0 \, .\label{EOM-Upsilon}
\end{align}
\end{subequations}
These equations govern the temporal evolution of Bondi mass and Bondi angular momentum at null infinity.

In summary, the solution space is characterized by six codimension-one functions subject to two constraints. Therefore, the solution space is parametrized by four $(6-2)$ unconstrained functions.

In solutions that involve free parameters, we often inquire whether these parameters correspond to Noether charges associated with certain symmetries. For example, the Schwarzschild solution has a free mass parameter, which is the conserved charge related to time translation symmetry. In our case, our solution for Einstein’s equation \eqref{metric}-\eqref{metric-components} includes four free functions. We explore whether these functions can be interpreted as charges for some symmetries. Indeed, this is the case, and the symmetry generators are expressed as follows
\begin{equation}\label{null-bondary-sym-gen}
    \xi = T\, \lambda\, \partial_{v}+\left[Y-\mathcal{U}\, \lambda \, T+\frac{\partial_{\phi}(T\lambda)}{\lambda\, \mathcal{R}}\right]\partial_{\phi} + \xi^r \partial_r\, ,
\end{equation}
with
\begin{equation}
    \begin{split}
        \xi^r = & \frac{1}{\eta \, \lambda} \Biggl[ Z -T \, \lambda \, [\partial_{v} \Omega-\partial_{\phi}(\mathcal{U}\, \Omega)]-\partial_{\phi}(\Omega Y)\\
        &- \frac{1}{\eta }\partial_{\phi}\left(\frac{\eta\,\partial_{\phi}(T\lambda)}{\lambda}\right)\Biggr]-\frac{{\Upsilon}-\Omega \, \partial_{\phi}\Pi }{2\, \eta\, \lambda^2\, R} \, \partial_{\phi}(T\lambda)\\
        &- \frac{r }{2}\, \left[  W +  T\, \lambda \,  (\partial_v \Pi-\mathcal{U}\partial_{\phi}\Pi) -2\, e^{\Pi/2} Z +Y\, \, \partial_{\phi}\Pi\right] \, .
    \end{split}
\end{equation}
Given our solution incorporates four free codimension-one functions, we anticipated an equivalent number of free codimension-one functions for the corresponding symmetry generators. The vector field mentioned above meets this expectation, being parametrized by four functions of $v,\phi$: supertranslations $T(v,\phi)$ and $Z(v,\phi)$, superscaling $W(v,\phi)$, and superrotation $Y(v,\phi)$. 
\section{Solution phase space without constraints}\label{sec:solution-phase-space-without-const}
In this section, our goal is to construct a solution phase space free of any constraints. In other words, we solve constraint equation \eqref{M-Upsilon-EoM}. Leveraging the fact that all solutions of Einstein’s equations (without a cosmological constant) in three dimensions are locally flat, we anticipate a coordinate transformation that can map the metric \eqref{metric} to Minkowski spacetime and vice versa. This section is dedicated to introducing such a coordinate transformation.

Let us start from $3d$ Minkowski spacetime
\begin{equation}\label{background-vac}
    \d{}s^2=2 \d{}w\,  \d{}\rho + \rho^2 \d{}\psi^2\, ,
\end{equation}
where $w$ and $\rho$ are two null coordinates and $\psi$ is a spatial coordinate \footnote{If we restrict $\psi$ to be a periodic coordinate, $\psi \sim \psi+2\pi$, then the line element will be described an orbifold of Minkowski spacetime.}. Now let us consider the following finite coordinate transformation, $\{w, \rho, \psi\} \to \{v, r, \phi\}$,
\begin{equation}\label{coordinate-trans-mink}
            \tcbset{fonttitle=\scriptsize}
            \tcboxmath[colback=white,colframe=gray]{\begin{split}
            \rho & = \mathcal{R}_1+r\, \mathcal{R}_0\, ,\\
            w & =\mathcal{T}+\frac{1}{2}\left(\frac{\partial_{\phi}\mathcal{T}}{\partial_{\phi}\mathcal{Y}}\right)^2  \frac{1}{\mathcal{R}_1+r\, \mathcal{R}_0}\, , \\
            \psi & = \mathcal{Y}+\frac{\partial_{\phi} \mathcal{T}}{\partial_{\phi}\mathcal{Y}} \frac{1}{\mathcal{R}_1+r\, \mathcal{R}_0} \, ,
         \end{split}}
\end{equation}
where $\mathcal{R}_0(v,\phi)$, $\mathcal{R}_1(v,\phi)$, $\mathcal{T}(v,\phi)$, and $\mathcal{Y}(v,\phi)$ are four free functions \footnote{One might question whether these transformations become singular when the denominator vanishes. However, if we assume these transformations to have an infinitesimal version connected to the identity, then the denominator will not vanish. This assumption ensures the regularity of the transformations.}. Upon performing the coordinate transformation, it can be verified that the line element given by equation \eqref{background-vac} transforms into the line element \eqref{metric}-\eqref{metric-components}. The six original functions $\{\eta,  \mathcal{U}, \Omega, \lambda, \mathcal{M}, \Upsilon \}$ in terms of four geometric functions $\{\mathcal{R}_0, \mathcal{R}_1, \mathcal{T}, \mathcal{Y}\}$ are given by
\begin{subequations}\label{full-on-shell}
\begin{align}
        &\Omega =  \partial_{\phi}\mathcal{Y}\, \mathcal{R}_1+\partial_{\phi}\left(\frac{\partial_{\phi}\mathcal{T}}{\partial_{\phi}\mathcal{Y}}\right)\, ,\qquad \\
        &\eta =  \mathcal{R}_0 \left(\partial_{v}\mathcal{T}-\frac{\partial_{v}\mathcal{Y}\, \partial_{\phi}\mathcal{T}}{\partial_{\phi}\mathcal{Y}}\right)\, ,\\
        & \lambda =  \frac{(\partial_{\phi}\mathcal{Y})^2}{\partial_{v}\mathcal{T}\partial_{\phi}\mathcal{Y}-\partial_{v}\mathcal{Y}\partial_{\phi}\mathcal{T}}\, , \quad \\
        &\mathcal{U} =  \frac{\partial_{v}\mathcal{Y}}{\partial_{\phi}\mathcal{Y}} \, , \\
        &{\mathcal{M}} =  -2\, \text{Sch}[\mathcal{Y}; \phi]\, , \label{M-sol}\\
        & {\Upsilon} =  -2\partial_{\phi}\mathcal{Y}\left[\partial_{\phi}\mathcal{Y}\partial_{\phi}\left(\frac{\partial_{\phi}^2 \mathcal{T}}{(\partial_{\phi}\mathcal{Y})^3}\right)-\partial_{\phi}\mathcal{T}\partial_{\phi}\left(\frac{\partial_{\phi}^2 \mathcal{Y}}{(\partial_{\phi}\mathcal{Y})^3}\right)\right]\, .
\end{align}
\end{subequations}
Interestingly, a direct computation confirms that the constraint equations, as given by \eqref{M-Upsilon-EoM}, are indeed satisfied when we apply equation \eqref{full-on-shell}.  It’s noteworthy that by solving these constraint equations, we are left with a usual unconstrained system. 

In summary, thanks to the finite coordinate transformation \eqref{coordinate-trans-mink}, we discovered a fully on-shell solution space for three-dimensional Einstein’s gravity, characterized by four arbitrary codimension-one functions: $\{\mathcal{R}_0(v,\phi), \mathcal{R}_1(v,\phi), \mathcal{T}(v,\phi), \mathcal{Y}(v,\phi)\}$. By construction, solution \eqref{metric} with \eqref{metric-components}-\eqref{full-on-shell} is \textit{locally} diffeomorphic to $3d$ Minkowski spacetimes and these four functions parameterize the Minkowski coadjoint orbit.

\section{Darboux Basis}\label{sec:Darboux Basis}
In this section, we consider the symplectic structure of theory \eqref{action}. We compute the symplectic potential, symplectic form, surface charge variation, and the corresponding charge algebra. 

By taking the first-order variation of action \eqref{action}, we get
\begin{equation}
    \delta S=\int \d{}^{3} x\, (E_{\mu\nu}\, \delta g^{\mu\nu}+\partial_{\mu}\Theta^\mu)\, ,
\end{equation}
where $E_{\mu\nu}=0$ gives Einstein's field equation, $R_{\mu\nu}=0$. The boundary term, which arises from integration by parts, is referred to as the pre-symplectic potential \cite{lee:1990nz, Iyer:1994ys, Wald:1999wa}. Its explicit form is as follows
\begin{equation}\label{Theta-mu-generic}
  \Theta^\mu[ g ; \delta g]:=\Theta_{_{\text{LW}}}^\mu [ g ; \delta g] + \nabla_{\nu} Y^{\mu \nu}[ g ; \delta g]+\delta  \mathcal{L}_{\mathcal{B}}^{\mu}[ g ] \, ,
\end{equation}
where $\Theta^{\mu}_{_{\text{LW}}} [g; \delta g]$ is the Lee-Wald symplectic potential for pure gravity,  
\begin{equation}\label{Theta-LW}
    \Theta^{\mu}_{_{\text{LW}}} [g; \delta g]:=\frac{\sqrt{-g}}{8 \pi G} \nabla^{[\alpha} \left( g^{\mu ] \beta} \delta g_{\alpha \beta} \right)\, ,
\end{equation}
and $Y^{\mu\nu}$ is a skew-symmetric tensor constructed out of metric and its variation. We refer to this tensor as $Y-$freedom and fix it as follows
\begin{equation}\label{Y-r-removing}
    Y^{\mu\nu}[\delta g; g]=\frac{1}{8 \pi G}\left(2\delta \sqrt{-g}\, n^{[\mu}l^{\nu]}+3\sqrt{-g}\, \delta n^{[\mu}l^{\nu]} \right)\, ,
\end{equation}
where $l^\mu$ and $n^\nu$ are two null vector fields
\begin{equation}\label{null-basis}
    l^{\mu}\partial_{\mu} =\partial_{v}+\frac{V}{2\eta}\partial_{r}-U\partial_{\phi}\, , \qquad n^{\mu}\partial_{\mu}=-\frac{1}{\eta}\partial_{r}\, ,
\end{equation}
such that they are normalized as $n \cdot l =-1$. We fixed $Y$-freedom \eqref{Y-r-removing} upon the requirement that symplectic potential and hence the surface charge variations are finite and $r$ independent. The boundary Lagrangian will be fixed later, requiring a well-defined variational principle. 

With this $Y$-term, the symplectic potential on arbitrary constant $r$ surfaces $\bTh:=  \int \Theta^\mu \d{}^2 x_\mu$ takes the form \cite{Adami:2022ktn} \footnote{It’s crucial to reemphasize that due to the $Y$-freedom \eqref{Y-r-removing}, the symplectic potential is $r$-independent. Consequently, one can interpret $\bTh$ as the symplectic potential over any $r$-constant hypersurface.}
\begin{equation}\label{symp-pot-partially-onshell}
\begin{split}
    \bTh=&\int \d{}v\,  \d{}\phi\,[\lambda^{-1}\delta{\mathcal{M}}-{\Upsilon}\, \delta\mathcal{U}+\partial_{v}(\Omega\, \delta\Pi)+\delta \mathcal{L}_{\mathcal{B}}]\, ,
\end{split}
\end{equation}
where $\mathcal{L}_{\mathcal{B}}:=\mathcal{L}_{\mathcal{B}}^{r}$. The first two terms in \eqref{symp-pot-partially-onshell} are referred to as the hydrodynamic or codimension-one part of the symplectic potential. This terminology is used because these terms are components of a conserved energy-momentum tensor (refer to section \ref{sec:Schwarzian-action}). The third term is designated as a corner or codimension-two part, given its nature as a total derivative term.

The hydrodynamic part of symplectic potential\eqref{symp-pot-partially-onshell} indicates that $\lambda$ and $\mathcal{U}$ are chemical potentials associated with the hydrodynamic variables $\mathcal{M}$ and $\Upsilon$ respectively.  However, it’s important to remember that they are not independent, but rather, they are governed by the constraint equations \eqref{M-Upsilon-EoM}. Conversely, the corner component involves quantities that are independent, and as a result, $\Omega$ and $\Pi$ form a Heisenberg conjugate pair. In this regard, the expression \eqref{symp-pot-partially-onshell} represents a "partially on-shell" expression for the symplectic potential.

Now, we are interested in computing the fully on-shell symplectic potential. By employing equation \eqref{full-on-shell}, we obtain \footnote{In this section, we implicitly assumed $\phi$ is a $2\pi$-periodic coordinate on a circle, $\phi \sim \phi+2\pi$.  Additionally, we assumed functions incorporated in the metric to exhibit periodicity (e.g. $\mathcal{T}(v,\phi+2\pi)=\mathcal{T}(v,\phi)$). Investigating the role of these total $\phi$-derivative terms would be an intriguing avenue for further study.}
\begin{equation}\label{Symp-Pot}
    \begin{split}
        \bTh=& \frac{1}{16 \pi G}\int \d{}v\,  \d{}\phi\, \partial_{v}(-\mathcal{J}\delta \mathcal{Y}+\Omega\, \delta\Pi)+\int \d{}^2 x_\mu\, \delta  \mathcal{L}_{\mathcal{B}}^{\mu},
    \end{split}
\end{equation}
where we have defined $\mathcal{J}:=\frac{{\Upsilon}}{\partial_{\phi}\mathcal{Y}}$. By taking the $v$-integral we find
\begin{equation}\label{Symp-Pot}
\bTh=  \frac{1}{16 \pi G}\oint \d{}\phi\,(-\mathcal{J}\delta \mathcal{Y}+\Omega\, \delta\Pi)+\int \d{}^2 x_\mu\, \delta  \mathcal{L}_{\mathcal{B}}^{\mu}\, .
\end{equation}
This symplectic potential exhibits some fascinating characteristics: 1) It is entirely a corner term. 2) It has Darboux form in the corner: $\{\Omega, \Pi\}$ and $\{\mathcal{J},\mathcal{Y}\}$ are two Heisenberg pairs.  3) For the establishment of a well-defined action principle, there is no requirement for any boundary Lagrangian $\mathcal{L}_{\mathcal{B}}^{\mu}=0$. 

The first characteristic underscores the holographic and topological nature of three-dimensional Einstein’s gravity. The second characteristic is anticipated, given that in terms of on-shell quantities \eqref{full-on-shell}, the constraint equations \eqref{M-Upsilon-EoM} are satisfied. Consequently, the theory behaves as a conventional unconstrained theory, and the symplectic pairs in the symplectic potentials emerge as Heisenberg pairs. The third characteristic reveals that, contrary to the common approach of ensuring a well-defined action principle that necessitates the addition of boundary terms, we do not require any boundary terms in our case.

Finally, the corresponding symplectic two-form is given by
\begin{equation}\label{Symp-form}
            \tcbset{fonttitle=\scriptsize}
            \tcboxmath[colback=white,colframe=gray]{
            \bO = \frac{1}{16 \pi G}\oint \d{}\phi\,(\delta \mathcal{Y} \wedge \delta\mathcal{J}+\delta \Omega \wedge\, \delta\Pi)\, .
            }
\end{equation}
Having obtained the symplectic form, we are now in a position to compute the surface charge associated with asymptotic symmetries as per equation \eqref{null-bondary-sym-gen}. To accomplish this, it is sufficient to contract one of the variations in the symplectic form with $\xi$ \cite{lee:1990nz, Iyer:1994ys, Wald:1999wa},
\begin{equation}
    \begin{split}
        \delta Q_\xi &:=\bO [\delta g ,\delta_\xi g; g]  \\
        &= \frac{1}{16 \pi G}\oint \d{}\phi\Big[\delta_{\xi} \mathcal{Y}  \delta\mathcal{J}-\delta \mathcal{Y}  \delta_{\xi}\mathcal{J}\\
        &\hspace{2.2 cm}+\delta_{\xi} \Omega \, \delta\Pi-\delta \Omega \, \delta_{\xi}\Pi\Big]\, .
    \end{split}
\end{equation}
The above form of the surface charge suggests defining new symmetry generators which we refer to as the Darboux generators
\begin{equation}\label{symmetry-generator-Darboux}
    \tilde{Z} :=\delta_{\xi}\Omega \, , \quad \tilde{W}:=-\delta_{\xi}\Pi \, , \quad \tilde{Y}:= \delta_{\xi}\mathcal{Y}\, , \quad \tilde{T}:= -\delta_{\xi}\mathcal{J} \, .
\end{equation}
In terms of Darboux generators, the surface charge variation becomes
\begin{equation}\label{surface-charge-Darboux}
    \begin{split}
        \slashed{\delta} Q_\xi &:=\bO [\delta g ,\delta_\xi g; g]  \\
        &= \frac{1}{16 \pi G}\oint \d{}\phi\left[\tilde{Y} \delta\mathcal{J}+\tilde{T}\delta \mathcal{Y}  +\tilde{Z} \, \delta\Pi+\tilde{W}\delta \Omega \, \right]\, .
    \end{split}
\end{equation}
By assuming $\delta \tilde{Z}=\delta \tilde{W}=\delta \tilde{Y}=\delta \tilde{T}=0$, it becomes evident that the charge variation \eqref{surface-charge-Darboux} is manifestly integrable
\begin{equation}\label{surface charge-Darboux-2}
            \tcbset{fonttitle=\scriptsize}
            \tcboxmath[colback=white,colframe=gray]{
            Q_\xi =\frac{1}{16\pi G} \oint \d \phi \left(
            \tilde{W}\, \Omega +\tilde{Z} \, \Pi+\tilde{Y} \, \mathcal{J}  + \tilde{T} \, \mathcal{Y} \right)\, .
            }
\end{equation}
We refer to the surface charge $\{\Omega, \Pi, \mathcal{J}, \mathcal{Y}\}$ as the "Darboux Soft Hair". This terminology arises from \cite{Hawking:2016msc, Haco:2018ske}. It is important to emphasize that the term ‘soft hair’ typically refers to the surface charges over the horizon. However, in this context, we apply this terminology to our surface charge \eqref{surface charge-Darboux-2}, which is defined over any $r$-constant hypersurface.

Having the Darboux symmetry generators \eqref{symmetry-generator-Darboux} and the corresponding surface charges \eqref{surface-charge-Darboux} we can compute their associated algebras.

\textbf{Boundary symmetry algebra.}
 Using the adjusted Lie bracket \cite{Barnich:2010eb, Compere:2015knw} we find the following algebra among Darboux symmetry generators
\begin{equation}\label{3d-NBS-KV-algebra}
    [\xi(  \tilde{T}_1, \tilde{Z}_1, \tilde{W}_1, \tilde{Y}_1), \xi( \tilde{T}_2, \tilde{Z}_2,  \tilde{W}_2, \tilde{Y}_2)]_{_{\text{adj.}}}=\xi(0, 0, 0, 0)\, .
\end{equation}
This is the direct sum of four copies of $U(1)$ algebra.

\textbf{Surface charge algebra.}
The charge algebra in the Darboux slicing is given by
\begin{equation}\label{3d-BT-Bracket-02''}
    \left\{Q_{\xi_{1}},Q_{\xi_{2}}\right\}_{{{\text{\tiny{MB}}}}} =\, Q_{[\xi_{1},\xi_{2}]_{{\text{adj.}}}}+K_{\xi_1,\xi_2}\, ,
\end{equation}
where $K_{\xi_1,\xi_2}$ admits two Heisenberg central extension terms
\begin{equation}
   \begin{split}
        K_{\xi_1,\xi_2}=\frac{1}{16\pi G} \oint \d \phi \, (\tilde{Z}_{2}\tilde{W}_{1}-\tilde{Z}_{1}\tilde{W}_{2}+\tilde{Y}_{2}\tilde{T}_{1}-\tilde{Y}_{1}\tilde{T}_{2})\, .
   \end{split}
\end{equation}
The explicit form of the charge algebra is two copies of the Heisenberg algebra
\begin{equation}
            \tcbset{fonttitle=\scriptsize}
            \tcboxmath[colback=white,colframe=gray]{\begin{split}
            &\{\Omega(v,\phi),\Pi(v,\phi')\} = 16\pi G\ \delta\left(\phi-\phi'\right)\, ,  \\
            &\{\mathcal{Y}(v,\phi),\mathcal{J}(v,\phi')\}=16\pi G\ \delta\left(\phi-\phi'\right)\, .
         \end{split}}
\end{equation}

\textbf{Isometry.} The $3d$ Minkowski spacetime \eqref{background-vac} is a maximally symmetric solution and accommodates six isometries (refer to appendix \ref{sec:isometry}).  Given that our solution space is locally diffeomorphic to the three-dimensional Minkowski spacetime, it will inherently possess six local isometries. 

One of the remarkable features of the Darboux basis is its ability to display the isometries of the solution manifestly. More precisely, the isometries are characterized by algebraic equations expressed in terms of the Darboux generators \eqref{symmetry-generator-Darboux}. For instance, when $\tilde{W}=\tilde{Z}=\tilde{Y}=\tilde{T}=0$ and $\tilde{Y}=1, \tilde{T}= \tilde{W}=\tilde{Z}=0$ we find two Killing vector fields which respectively correspond to $\partial_w$ and $\partial_{\psi}$ in the original coordinate \eqref{background-vac} (for the explicit form of these vector fields see appendix \ref{sec:isometry}). It is noteworthy that, in contrast to the algebraic method employed here, other bases often necessitate the resolution of differential equations, such as Hill’s equation, to ascertain isometries (refer to \cite{Sheikh-Jabbari:2014nya, Sheikh-Jabbari:2016unm} for further details).
\section{Two Schwarzian actions}\label{sec:Schwarzian-action}
In this section, we demonstrate that the requirement of a well-defined action principle with Dirichlet boundary conditions leads to two Schwarzian actions for boundary theory.

In \cite{Adami:2024rkr}, an intrinsic stress tensor and current were obtained for three-dimensional Einstein gravity in asymptotically flat spacetimes (see also \cite{Hartong:2015usd}, and for a related topic \cite{Campoleoni:2018ltl}). Additionally, \cite{Adami:2024rkr} demonstrated that the requirement of a well-defined action principle results in a boundary action involving one Schwarzian action. Intriguingly, we show here that the on-shell boundary action yields two Schwarzian actions at null infinity.

First, we will provide a concise review of the intrinsic construction as outlined in \cite{Adami:2024rkr}. The conformal induced metric at null infinity is defined as follows
\begin{equation}
    \gamma_{a b}=k_{a} k_{b}, \qquad  k_{a} \d x^a:= \d \phi + \mathcal{U} \d v \, .
\end{equation}
As it is clear from the form of the induced metric, it is degenerate and its kernel is given by 
\begin{equation}\label{l-def}
    l^{a}\partial_{a}:= \lambda(\partial_{v}-\mathcal{U}\partial_{\phi})\, .
\end{equation} 
The one-form dual of the kernel vector, the Ehresmann connection, is defined as follows
\begin{equation}\label{n-def}
n_{a}\d{}x^a=-\lambda^{-1} \d{}v\, , \qquad l^{a}n_{a}=-1\, .
\end{equation} 
The triplet $\{\gamma_{ab}, l^{a}, n_a\}$ provides a ruled Carrollian structure \cite{Duval:2014uoa, Duval:2014uva, Duval:2014lpa, Henneaux:1979vn, Henneaux:2021yzg, Ciambelli:2019lap,deBoer:2021jej, deBoer:2017ing} at null infinity.

Using this structure one can define a projection operator 
\begin{equation}\label{Proj-def}
P^{a}{}_{b}:=\delta_{b}^{a}+n_{b}l^{a}\, , \qquad P^{a}{}_{b}l^{b}= P^{a}{}_{b}n_{a}=0\, ,
\end{equation} 
and subsequently a partial inverse, $h^{ac}\gamma_{cb}=P^{a}{}_{b}$, with the following explicit form
\begin{equation}\label{h-def}
h^{ab}=k^{a}k^{b}\, , \qquad k^{a}\partial_{a}:=\partial_{\phi}\, .
\end{equation}
Finally, we augment the intrinsic geometry of null infinity by introducing the following torsion-full connection
\begin{equation}\label{connection}
    \Gamma^{c}_{ab}= \frac{1}{2}h^{cd}\left(\partial_{a}\gamma_{bd}+\partial_{b}\gamma_{ad}-\partial_{d}\gamma_{ab}\right)+h^{cd}\,K_{da}n_{b}+l^c S_{ab}\, ,
\end{equation}
where 
\begin{equation}\label{S-ab}
    K_{ab}:= \frac{1}{2}\mathcal{L}_{l}\gamma_{ab}, \quad S_{ab}:=2\partial_{c}l^{c}\, n_{a}n_{b}-3\,\partial_{(a}n_{b)}{- n_{[a} \mathcal{L}_l n_{b]}}.
\end{equation}
Utilizing this connection, we define the expansions of the vector fields $l^a, k^a$, namely $\theta_l, \theta_k$, 
\begin{equation}\label{expansions}
    \theta_l:=D_{a}l^a=-2\, \lambda \,  \mathcal{U}' \, , \qquad \theta_k:=D_{a}k^{a} ={-2\,  \lambda^{-1} \lambda'} \, .
\end{equation}
Now, we rewrite the symplectic potential in terms of hydrodynamics variables
\begin{equation}\label{SP-01}
        \begin{split}
            \bTh &= {\frac{1}{16\pi G}}\int \d{}^2 x \sqrt{\gamma} \left( \text{T}^{ab} \delta \gamma_{ab}+ P^{a} \delta n_{a}\right)\\
            &+\frac{1}{16 \pi G} \int  \d{}^2 x \, \partial_v\left( \Omega \, \delta \Pi \right)  \\
            &+ \delta \int \d{}^2 x \left(L_{\mathcal{B}}+ \frac{\mathcal{M}}{16\pi\, G \sigma'}+\frac{\mathrm{Sch}[\sigma;\phi]}{8\pi G \sigma'} \right)\, ,
        \end{split}
\end{equation}
where the explicit forms of the stress tensor and current are 
\begin{equation}\label{cons-EMT-total}
    \begin{split}
       \text{T}^{a}{}_{b}&:= \Big(\mathcal{M}- k^c \partial_c \theta_{k}-\frac{1}{4}\theta_{k}^{2}\Big)\,  k^a k_b  -\Upsilon \, l^{a}k_{b}\, ,\\
       \text{P}^{a}&:=\Big({\cal M}- k^c \partial_c \theta_{k}-\frac{1}{4}\theta_{k}^{2}\Big) \, l^a-\left(k^c \partial_c \theta_l-l^c \partial_c \theta_k\right) \, k^a\, .
    \end{split}
\end{equation}
As demonstrated in \cite{Adami:2024rkr}, these hydrodynamic variables are conserved, i.e., $D_{a}P^a=0$ and $D_{a}\text{T}^{a}{}_{b}=0$. In essence, they reproduce constraint equations \eqref{M-Upsilon-EoM}. On the other hand, if we employ on-shell variables \eqref{full-on-shell}, these hydrodynamic equations are satisfied automatically.

Now, at this stage, we require a well-defined action principle with the Dirichlet boundary condition. To achieve this, we must choose $\mathcal{L}_{\mathcal{B}}$ such that it causes the total variation term (the last line) in \eqref{SP-01} to vanish. Then, we obtain
\begin{equation}\label{one-Schwarzians}
            S_{\mathcal{B}}= -\frac{1}{8\pi\, G}\int \d v \int_0^{2\pi} \frac{\d \phi}{\sigma'} \left(\frac{\mathcal{M}}{2}+\, \mathrm{Sch}[\sigma;\phi]\right).
\end{equation}
This refers to the single-Schwarzian boundary action that was derived in \cite{Adami:2024rkr}. Now, by utilizing the on-shell value of $\mathcal{M}$ \eqref{M-sol}, we identify two Schwarzians for the boundary action
\begin{equation}\label{Two-Schwarzians}
            \hspace{-.17 cm}\tcbset{fonttitle=\scriptsize}
            \tcboxmath[colback=white,colframe=gray]{\hspace{-.25 cm}
            S_{\mathcal{B}}
    = \frac{1}{8\pi\, G}\int \d v \int_0^{2\pi} \frac{\d \phi}{\sigma'} \left(\text{Sch}[\mathcal{Y}; \phi]-\, \mathrm{Sch}[\sigma;\phi]\right).
            \hspace{-.25 cm}}
\end{equation}
This represents one of our key results. In the subsequent section, we will illustrate that these two Schwarzians are related to the Virasoro algebras at null infinity.
\section{Various Lie-algebra}\label{sec:Various Lie-algebra}
This section explores a variety of algebras that are derived from choosing various field-redefinitions in solution space, a process known as change of slicing \cite{Adami:2020ugu, Adami:2021sko, Adami:2021nnf, Adami:2022ktn, Taghiloo:2022kmh}.

\textbf{Two copies of current algebra.} In order to realize a variety of Lie algebras, we introduce the following currents
\begin{equation}\label{def-j+-}
    \begin{split}
        &J_1^\pm := \frac{1}{16\pi G} \left( \Omega \mp 2 G k_1 \, \partial_\phi \Pi \right)\, ,\\
        &J_2^\pm := \frac{1}{16\pi G} \left( \mathcal{Y} \mp 2 G k_2 \, \partial_\phi \mathcal{J} \right)\, ,
    \end{split}
\end{equation}
where $k_1$ and $k_2$ are two arbitrary constants. Next, we establish the relationship among the chemical potentials
\begin{equation}
    \begin{split}
        &\tilde{W}=  \epsilon_1^+ + \epsilon_1^-  \, , \qquad \tilde{Z}= 2 G k_1\, \partial_\phi \left( \epsilon_1^+ - \epsilon_1^- \right) \, ,\\
        &\tilde{T}=  \epsilon_2^+ + \epsilon_2^-  \, , \qquad \tilde{Y}= 2 G k_2\, \partial_\phi \left( \epsilon_2^+ - \epsilon_2^- \right) \, .
    \end{split}
\end{equation}
Therefore, the charge variation in this basis is given by
\begin{equation}
\delta Q_\xi =  \sum_{i=1,2}\,\int_0^{2\pi} \d \phi \left( \epsilon_i^+ \delta J_i^+ + \epsilon_i^- \delta J_i^-\right) \, .
\end{equation}
Consequently, the charge algebra can be expressed as follows
\begin{equation}\label{Currents+Witt}
\begin{split}
&\{ J_i^\pm (v,\phi), J_j^\pm (v,\phi') \}= \pm \frac{k_i}{4\pi} \delta_{ij}\, \partial_{\phi} {\delta} (\phi-\phi')\, ,\\
& \{ J_i^\pm (v,\phi), J_j^\mp (v,\phi') \}=0\, .
\end{split}
\end{equation}

\textbf{Four copies of Virasoro algebras.} These currents enable the incorporation of Virasoro generators
\begin{equation}\label{L+-}
    \begin{split}
        &L_i^\pm (v,\phi):=  \frac{2\pi}{k_i} \left[ J_i^\pm( v,\pm \phi)\right]^2 +{\beta_i^\pm} \partial_\phi J_i^\pm(v,\pm \phi)\, ,
    \end{split}
\end{equation}
here, $\beta_i^{\pm}$ represents four arbitrary constants, with $i=1,2$, are four arbitrary constants. The corresponding chemical potentials $\chi_i^{\pm}$ are given in terms of previous chemical potentials as follows
\begin{equation}
    \epsilon_i^\pm(v,\phi) = \left(\frac{4\pi}{k_i}   J_i^\pm(v,\phi) \mp {\beta_i^\pm} \partial_\phi \right) \chi_i^\pm( v,\pm \phi)\, .
\end{equation}
Then the charge variation is
\begin{equation}
   \delta Q_{\xi}=\sum_{i=1,2} \int_0^{2\pi} \d \phi \, (\chi_i^{+} \delta L_i^+ +\chi_i^{-} \delta L_i^-)\, .
\end{equation}
The variations of Virasoro generators $L_i^\pm$ are hence
\begin{equation}
    \delta_\xi L_i^\pm = \chi_i^\pm \partial_\phi L_i^\pm +2 L_i^\pm \partial_\phi \chi_i^\pm - \frac{k_i}{4\pi} {(\beta_i^\pm)^2}\partial_\phi^3 \chi_i^\pm \, .
\end{equation}
Ultimately, we derive the following surface charge algebra
\begin{subequations}
    \begin{align}
    \{L_i^\pm (v,\phi), L_j^\pm (v,\phi')\}=& \delta_{ij}\Big( L_i^\pm (v,\phi')\,\partial _\phi-L_i^\pm (v,\phi)\,\partial_{\phi'}\nonumber \\
    & \hspace{1 cm} +\frac{k_i}{4\pi} {(\beta_i^\pm)^2} \partial_{\phi'}^3 \Big) {\delta} (\phi-\phi')\, , \label{LpmLpm}\\
    \{L_i^\pm (v,\phi), L_j^\mp (v,\phi')\}=& 0\, .
    \end{align}
\end{subequations}
This represents a direct sum of four Virasoro algebras, each with an arbitrary central charge defined as $c_i^\pm=6 k_i (\beta_i^\pm)^2$.
It is noteworthy that the emergence of two Schwarzian boundary actions at null infinity \eqref{Two-Schwarzians}  is a direct consequence of these Virasoro algebras.

\textbf{Two copies of BMS$_3$.}
As another intriguing example, we construct two copies of BMS$_3$ algebras. To do this, we initially construct the following two currents on an off-diagonal basis,
\begin{equation}\label{def-JK}
    J_i:= J_i^+ + J_i^-,\qquad K_i:= J_i^+- J_i^-\, .
\end{equation}
From these, we can define the BMS generators as follows:
\begin{equation}\label{L&M}
    L_i= \frac{{2}\pi}{k_i} J_i\, K_i + {\beta}_i\partial_\phi K_i+ {\alpha}_i\, \partial_\phi J_i\, , \quad M_i= \frac{\pi}{k_i} J_i^2 + {\beta_i}\partial_\phi J_i \, .
\end{equation}
The charge variation is then given by
\begin{equation}
   \delta Q_{\xi}= \sum_{i=1,2}\int_0^{2\pi} \d \phi \, (\epsilon^{\text{\tiny{L}}}_i \delta L_i +\epsilon^{\text{\tiny{M}}}_i \delta M_i )\, , 
\end{equation}
here chemical potentials $\epsilon^{\text{\tiny{L}}}_i$ and $\epsilon^{\text{\tiny{M}}}_i$ are defined as follows
\begin{equation}
\epsilon_i^\pm = \frac{{2}\pi}{k_i} \left[ (K_i \pm J_i ) \epsilon_i^{\text{\tiny{L}}} +  \epsilon^{\text{\tiny{M}}}_i J_i\right] - \beta_i \partial_\phi (\epsilon^{\text{\tiny{M}}}_i \pm \epsilon^{\text{\tiny{L}}}_i)-\alpha_i \partial_{\phi}\epsilon^{\text{\tiny{L}}}_i\, .
\end{equation}
The BMS fields transform under symmetry generators as
\begin{equation}
    \begin{split}
        \delta_\xi L_i=&\partial_\phi L_i \,  \epsilon^{\text{\tiny{L}}}_i + 2 L_i \partial_\phi \epsilon_i^{\text{\tiny{L}}} +\partial_\phi M_i \epsilon^{\text{\tiny{M}}}_i + 2 M_i \partial_\phi \epsilon^{\text{\tiny{M}}}_i \\
        &- \frac{k_i}{2\pi} {\beta_i^2} \partial_\phi^3 \epsilon^{\text{\tiny{M}}}_i - \frac{k_i}{\pi} \alpha_i \beta_i \partial_\phi^3 \epsilon^{\text{\tiny{L}}}_i\, ,\\
         \delta_\xi M_i=& \partial_\phi M_i \epsilon^{\text{\tiny{L}}}_i + 2 M_i \partial_\phi \epsilon^{\text{\tiny{L}}}_i - \frac{k_i}{2\pi} {\beta_i^2} \partial_\phi^3 \epsilon_i^{\text{\tiny{L}}} \, .
      \end{split}
\end{equation}
Finally, we arrive at the following algebra
\begin{equation}
    \begin{split}
    &\{M_i (v,\phi), M_j (v,\phi')\}=0 \, , \\
    &\{L_i (v,\phi), L_j (v,\phi')\}=\delta_{ij} \Big(L_i (v,\phi')\,\partial _\phi-L_i (v,\phi)\,\partial_{\phi'}\\
    &\hspace{4 cm}+ \frac{k_i}{\pi} {\alpha_i\beta_i} \partial_{\phi'}^3  \Big) {\delta} (\phi-\phi')\, , \\
    &\{L_i (v,\phi), M_j (v,\phi')\}=\delta_{ij} \Big(M_i (v,\phi')\,\partial _\phi-M_i (v,\phi)\,\partial_{\phi'} \\
    &\hspace{4 cm} + \frac{k_i}{2\pi} {\beta_i^2} \partial_{\phi'}^3 \Big) {\delta} (\phi-\phi')\, , \\
    \end{split}
\end{equation}
which is two copies of centrally extended BMS$_3$ algebra with central charges $c^{\text{\tiny{LM}}}_i=12k_i \beta_i^2, c^{\text{\tiny{LL}}}_i=24k_i \alpha_i\beta_i$.
\section{Summary, Discussion, and Outlook}\label{sec:Outlook}
In this section, we present a summary of our findings, engage in a discussion, and offer insights into future directions. 

\textbf{Summary.} We have developed a fully on-shell solution space for three-dimensional Einstein’s gravity, with a zero cosmological constant. This is achieved by employing a finite coordinate transformation on Minkowski spacetime. This constructed solution space encompasses four geometric functions at null infinity. Two of these functions account for the reparametrization of time and angle over the null infinity cylinder, while the remaining two represent radial supertranslation and radial superscaling.
 
 The symplectic form, when expressed in terms of geometric functions, takes a Darboux form at the corner. This is a direct result of having a fully on-shell phase space, in conjunction with the inherent topological characteristics of three-dimensional gravity. Consequently, the resulting surface charges form two copies of Heisenberg algebra. By implementing a variety of surface charge redefinitions (also known as change of slicing), we have been able to construct an array of charge algebras.
 
 We showed that the fully on-shell solution space in the Darboux basis admits a well-defined action principle without any boundary Lagrangian. On the other hand, in the hydrodynamics frame, a well-defined action principle with Dirichlet boundary conditions yields a boundary action with two Schwarzians.

\textbf{Discussion.} While there are numerous points to consider, we will only highlight a few key ones. The first point to note is that the construction of the boundary action for the Abelian $U(1)$ Chern-Simons theory is quite straightforward, as outlined in \cite{Kim:2023vbj}. This is primarily because its bulk solutions can generally be solved. Conversely, the construction of the boundary action for three-dimensional Einstein gravity presents a more complex challenge \cite{Kim:2023vbj}. This complexity arises from the fact that its equations cannot be generally solved, leaving us with constraint equations. Intriguingly, our approach enables us to solve these constraint equations, thereby allowing us to obtain the desired boundary action in a manner similar to the Abelian $U(1)$ Chern-Simons theory. We aim to address these problems in the upcoming works.

The second key point to emphasize is that the charge algebra is not invariant under field redefinition, meaning it depends on the slicing of the solution space. It’s important to note that the symplectic form, in contrast, is invariant under field redefinitions. It is widely accepted that the asymptotic symmetry group of asymptotically AdS and flat spacetimes consists of two copies of the Virasoro and BMS algebras, respectively. This suggests that the corresponding boundary theory should exhibit conformal and BMS invariance, respectively.

However, as we have noted, a critical point is that the concept of asymptotic algebra does not maintain invariance under field redefinition. In this paper, which primarily focuses on the asymptotically flat case, we have derived a range of algebras. Notably, we found four copies of the Virasoro algebra, which is typically linked with the AdS case, not the flat one. This implies that the assertion that the dual boundary theory of asymptotically flat spacetimes should be a BMS-field theory is not a field redefinition invariant statement. By utilizing various field redefinitions, one can construct simple algebras and then attempt to identify the corresponding dual theory.

Furthermore, it is crucial to emphasize that some central charges in the charge algebras are arbitrary. This implies that one has the ability to eliminate central charges through field redefinitions. This particular observation could potentially have profound implications for our comprehension of both gauge theories and gravitational theories.

\textbf{Outlook.} The key insight that enables us to construct a fully on-shell solution space is the realization that all solutions of three-dimensional Einstein’s gravity are either locally flat or (A)dS. In \cite{Adami:2022ktn}, a solution space comprising six functions and two constraints has been derived for both asymptotically flat and (A)dS spacetimes. In this paper, we solved the constraint equations in the flat case. Our forthcoming research is aimed at constructing the fully on-shell variant of this solution space for asymptotically AdS spacetimes.

While the solution space constructed in this paper is quite general, it is not maximal. The maximal solution spaces for asymptotically flat and AdS spacetimes are characterized by twelve codimension-one functions and six constraints, \cite{Grumiller:2017sjh, Grumiller:2016pqb}. Our future work will focus on constructing a fully on-shell (unconstrained) solution space with arbitrary six codimension-one functions, the corresponding Darboux basis, and the boundary action.


\begin{acknowledgments}
We would like to thank Hamed Adami, Hamid Reza Afshar, Luca Ciambelli, Marc Geiller, Mohammad Mehdi Sheikh-Jabbari, and Mohammad Hassan Vahidinia for discussions or comments on the manuscript. We would also like to thank Hossein Yavartanoo for long-term collaborations and discussions on related topics.
\end{acknowledgments}

\appendix

\section{Isometry}\label{sec:isometry}
The $3d$ Minkowski spacetime \eqref{background-vac} is a maximally symmetric spacetime and possess the following six Killing vectors 
\begin{equation}
    K=K^{\mu}\partial_{\mu}=K^{w}\partial_{w}+K^{\rho}\partial_{\rho}+K^{\psi}\partial_{\psi}\, ,
\end{equation}
where 
\begin{equation}       
    \begin{split}
        K^{w} & =(k_1+k_6\, w+k_3\, \psi+k_4\, w\, \psi+k_5\, \psi^2)\, ,\\
        K^{\rho} & =-2k_5 -k_6\, \rho-k_4\, \rho\, \psi\, ,\\
        K^{\psi} & =k_2+k_6\, \psi+\frac{k_3}{\rho}+k_4\frac{w}{\rho}+2k_5 \frac{\psi}{\rho}+\frac{k_4}{2}\psi^2\, ,
    \end{split}
\end{equation}
here $k_i$, $i=1,\cdots,6$ are six constants. It is important to note that we have not imposed any periodicity condition on the $\psi$-coordinate. Should such a periodicity condition be applied, only two Killing vector fields, namely, $\partial_w$ and $\partial_{\psi}$, would preserve the periodicity condition. These two Killing vector fields transform under coordinate transformations \eqref{coordinate-trans-mink} as follows
\begin{align}
    &\xi^{v}=\frac{\lambda}{\partial_{\phi}\mathcal{Y}}\, ,\\
    &\xi^{\phi}=-\frac{1}{\partial_{\phi}\mathcal{Y}}\left(\mathcal{U}\, \lambda+\frac{\lambda \chi_1}{{R}}\right)\, ,\\
    &\xi^{r}=\frac{-1}{\eta \partial_{\phi}\mathcal{Y}}\left[\mathcal{D}_{v}{R}-\partial_{\phi}\chi_1-\chi_1\left(\frac{{\Upsilon}-\Omega \, \partial_{\phi}\Pi}{2{R}}+\frac{\partial_{\phi}\eta}{\eta}-\lambda\, \chi_1\right)\right]\, ,
\end{align}
and
\begin{align}
    &\xi^{v}=-\frac{\lambda\, \partial_{\phi}\mathcal{T}}{(\partial_{\phi}\mathcal{Y})^2}\, , \\
    &\xi^{\phi}=\frac{\chi_2}{\partial_{\phi}\mathcal{Y}}+\mathcal{U}\frac{\lambda\, \partial_{\phi}\mathcal{T}}{(\partial_{\phi}\mathcal{Y})^2}\, , \\
    &\xi^{r}=\frac{-2}{\eta\, \lambda\, \partial_{\phi}\mathcal{Y}}\Biggl\{\chi_2\Bigg({\Upsilon}+2\partial_{\phi}{R}-2{R}\frac{\partial_{\phi}^2\mathcal{Y}}{\partial_{\phi}\mathcal{Y}}\Bigg)-\frac{2\lambda \partial_{\phi}\mathcal{T}}{\partial_{\phi}\mathcal{Y}}\times\nonumber\\
        &\Bigg[\mathcal{D}_{v}{R}+\frac{{\mathcal{M}}}{2\lambda}-\partial_{\phi}^{2}(\lambda^{-1})-\frac{\partial_{\phi}^2\mathcal{Y}}{\lambda \partial_{\phi}\mathcal{Y}}\Bigg(\frac{\partial_{\phi}\lambda}{\lambda}-\frac{\partial_{\phi}^2\mathcal{Y}}{2\partial_{\phi}\mathcal{Y}}\Bigg)\Bigg]\Biggr\}\, ,
\end{align}
where $\mathcal{D}_{v}{R}:=\partial_{v}R-\partial_{\phi}(\mathcal{U}\, R)$ and 
\begin{equation}
    \begin{split}
        &\chi_1:=\lambda^{-1}\partial_{\phi}\ln(\lambda^{-1}\partial_{\phi}\mathcal{Y}) \, ,\\
        &\chi_2:=1-\frac{\partial_{\phi}\mathcal{T}}{\partial_{\phi}\mathcal{Y}}\partial_{\phi}\ln\left(\frac{\lambda\, \partial_{\phi}\mathcal{T}}{(\partial_{\phi}\mathcal{Y})^2}\right)\frac{1}{R}\, .
    \end{split}
\end{equation}
\bibliography{references}

\end{document}